\documentclass[12pt]{article}
\usepackage{amssymb}
\usepackage{amsmath}
\usepackage{graphicx}
\usepackage[unicode,bookmarks,bookmarksopen,bookmarksopenlevel=2,colorlinks,linkcolor=blue,citecolor=green]{hyperref}

\input{tcilatex}
\begin{document}

\title{\textbf{Integrability of Exceptional Hydrodynamic Type Systems}}
\author{Maxim V. Pavlov \\
{\small Lebedev Physical Institute of Russian Academy of Sciences,}\\
[-3pt] {\small Leninskij Prospekt 53, 119991 Moscow, Russia }}
\date{\today }
\maketitle

\begin{abstract}
In this paper we consider non-diagonalisable hydrodynamic type systems
integrable by the Extended Hodograph Method. We restrict our consideration
to non-diagonalisable hydrodynamic reductions of the Mikhal\"{e}v equation.
We show that families of these hydrodynamic type systems are reducible to
the Heat hierarchy. Then we construct new particular explicit solutions for
the Mikhal\"{e}v equation.
\end{abstract}

\tableofcontents

\newpage

\bigskip \bigskip

\begin{flushright}
\textit{dedicated to the 80th birthday of S.P. Novikov}
\end{flushright}

\section{Introduction}

\label{sec-intro}

The theory of two-dimensional integrable quasilinear systems of first order
by the Tsarev Generalised Hodograph Method was established in \cite{Tsar}.
Such systems we call semi-Hamiltonian hydrodynamic type systems. They have
all pairwise distinct roots of characteristic polynomials of velocity
matrices\footnote{%
Everywhere below we shall write just \textquotedblleft
root\textquotedblright\ for simplicity.} and they are diagonalisable.

In this paper we deal with \textit{integrable} hydrodynamic type systems,
which have just one root. These systems are non-diagonalisable. However,
they have infinitely many hydrodynamic conservation laws, commuting flows
and particular solutions. Instead of Generalised Hodograph Method we apply
the Extended Hodograph Method presented in \cite{MaksEgor}.

In a general case even diagonalisable hydrodynamic type systems with
pairwise distinct roots are not semi-Hamiltonian, i.e. they are
non-integrable. Classification and integration of hydrodynamic type systems
with double roots, triple roots, etc. become much more complicated from
combinatorial point of view. For instance, in the two-component case we have
two different sub-cases: two distinct roots and a one common root; in the
three-component case we have three different sub-cases: three distinct
roots, a one double root and a one triple root; in the four-component case
we have five different sub-cases: four distinct roots, a one double root,
two double roots, a one triple root and a one quadruple root, etc. This is
one of many reasons: why integrable hydrodynamic type systems with multiple
roots were not investigated earlier.

Another important reason was absence of any interesting examples known in
the theory of integrable systems. Just very recently (see \cite{KK}), such a
family of hydrodynamic type systems was extracted from integrable
hydrodynamic chain (see \cite{maksjmp})%
\begin{equation}
c_{k,t}=c_{k+1,x}+c_{1}c_{k,x},\text{ \ }k=1,2,...  \label{hydrochain}
\end{equation}%
It was shown in \cite{maksjmp} that this chain possesses infinitely many
integrable diagonalisable hydrodynamic reductions with pairwise distinct
roots. The authors of \cite{KK} proved that this chain also possesses
infinitely many integrable hydrodynamic reductions with multiple roots. Then
the case with a single root was deeply investigated in \cite{KO}, where the
authors constructed a link to the Heat hierarchy. In this paper we continue
to consider hydrodynamic reductions with a single root. We apply the
Extended Hodograph Method for linearisation of these hydrodynamic type
systems to the Heat hierarchy, i.e.%
\begin{equation*}
u_{t}=u_{xx},\text{ \ }u_{y}=u_{xxx},\text{ \ }u_{z}=u_{xxxx},...,
\end{equation*}%
and we derive corresponding solutions for the remarkable Mikhal\"{e}v
equation (see \cite{mikh}).

The paper is organised as follows: in Section \ref{sec:chains} we discuss
integrable hydrodynamic chains (\ref{comflo}), which possess infinitely many
non-diagonalisable hydrodynamic reductions with multiple roots. In Section %
\ref{sec:ext} we apply the Extended Hodograph Method for linearisation of
multi-component hydrodynamic reductions with a single multiple root to the
remarkable Heat hierarchy. In Subsection \ref{subsec:hodo} we consider the
simplest two-component case integrable by the classical hodograph
transformation. In Subsection \ref{subsec:extri} we reduce two commuting
three-component hydrodynamic type systems with a single triple root to the
Heat equation together with its first higher commuting flow. In Subsection %
\ref{subsec:extfo} we reduce an arbitrary number of commuting
multi-component hydrodynamic type systems with a single quadruple root to
the Heat hierarchy. In Section \ref{sec:mikh} we show that the integrable
three-dimensional linearly degenerate Mikhal\"{e}v system simultaneously
possesses multi-component non-diagonalisable hydrodynamic reductions with a
single multiple root. In the three-component case we construct corresponding
particular solutions presented in explicit forms. Finally in Conclusion \ref%
{sec:final} we discuss existence of such non-diagonalisable hydrodynamic
reductions for other integrable hydrodynamic chains.

\section{The Integrable Hydrodynamic Chain}

\label{sec:chains}

Integrable hydrodynamic chain (\ref{hydrochain}) possesses infinitely many
higher commuting flows (see \cite{maksjmp})%
\begin{equation}
c_{k,t_{n+1}}=\overset{n}{\underset{m=0}{\sum }}a_{m}c_{k+n-m,x},\text{ }%
k=1,2,...,\text{ }n=0,1,2,...,  \label{comflo}
\end{equation}%
where all polynomial functions $a_{k}(\mathbf{c})$ can be found from the
expansion ($\lambda \rightarrow 0$)%
\begin{equation}
1+\overset{\infty }{\underset{m=1}{\sum }}a_{m}\lambda ^{m}=\exp \left( 
\overset{\infty }{\underset{m=1}{\sum }}c_{m}\lambda ^{m}\right) .
\label{aca}
\end{equation}%
This means that polynomial functions $a_{k}(\mathbf{c})$ are nothing but the
well-known Bell polynomials (up to appropriate factorial multipliers).

For instance,%
\begin{equation}
a_{0}=1,\text{ \ }a_{1}=c_{1},\text{ \ }a_{2}=c_{2}+\frac{1}{2}c_{1}^{2},%
\text{ \ }a_{3}=c_{3}+c_{1}c_{2}+\frac{1}{6}c_{1}^{3}.  \label{ac}
\end{equation}%
Corresponding commuting hydrodynamic chains are%
\begin{equation}
c_{k,t}=c_{k+1,x}+c_{1}c_{k,x},\text{ \ }c_{k,y}=c_{k+2,x}+c_{1}c_{k+1,x}+%
\left( c_{2}+\frac{1}{2}c_{1}^{2}\right) c_{k,x},  \label{raz}
\end{equation}%
\begin{equation}
c_{k,z}=c_{k+3,x}+c_{1}c_{k+2,x}+\left( c_{2}+\frac{1}{2}c_{1}^{2}\right)
c_{k+1,x}+\left( c_{3}+c_{1}c_{2}+\frac{1}{6}c_{1}^{3}\right) c_{k,x},
\label{dva}
\end{equation}%
where we denoted $x=t_{1},t=t_{2},y=t_{3},z=t_{4}$.

The differential of (\ref{aca}) leads to recursive consequences ($n=1,2,...$)%
\begin{equation*}
da_{n+1}=\overset{n}{\underset{m=0}{\sum }}a_{m}dc_{n+1-m}.
\end{equation*}%
Thus (see (\ref{comflo})),%
\begin{equation*}
c_{1,t_{n+1}}=\overset{n}{\underset{m=0}{\sum }}a_{m}c_{n+1-m,x}=a_{n+1,x}.
\end{equation*}%
For instance,%
\begin{equation}
c_{1,x}=a_{1,x},\text{ \ }c_{1,t}=a_{2,x},\text{ \ }c_{1,y}=a_{3,x},\text{ \ 
}c_{1,z}=a_{4,x}.  \label{cons}
\end{equation}%
Thus, (\ref{comflo}) can be written in the form%
\begin{equation}
c_{k,t_{n+1}}=c_{k+n,x}+\overset{n}{\underset{m=1}{\sum }}\Phi
_{m}c_{k+n-m,x},  \label{chaf}
\end{equation}%
where $\Phi $ is a potential function for conservation laws (\ref{cons}),
i.e. $a_{k}=\Phi _{k}$. Here $\Phi _{m}\equiv \partial \Phi /\partial t_{m}$.

\textbf{Remark}: Infinitely many conservation law densities $\sigma _{k}(%
\mathbf{c})$ can be found from the expansion ($\lambda \rightarrow 0$)%
\begin{equation}
1+\overset{\infty }{\underset{m=1}{\sum }}\sigma _{m}\lambda ^{m}=\exp
\left( -\overset{\infty }{\underset{m=1}{\sum }}c_{m}\lambda ^{m}\right) .
\label{conslo}
\end{equation}%
This means, that (cf. (\ref{aca}))%
\begin{equation*}
a_{1}+\sigma _{1}=0,\text{ \ }a_{2}+a_{1}\sigma _{1}+\sigma _{2}=0,\text{ \ }%
a_{3}+a_{2}\sigma _{1}+a_{1}\sigma _{2}+\sigma _{3}=0,...
\end{equation*}%
Taking into account (\ref{ac}), one can obtain again Bell polynomials (with
another choice of multipliers)%
\begin{equation*}
\sigma _{1}=-c_{1},\text{ \ }\sigma _{2}=-c_{2}+\frac{1}{2}c_{1}^{2},\text{
\ }\sigma _{3}=-c_{3}+c_{1}c_{2}-\frac{1}{6}c_{1}^{3},...
\end{equation*}%
Corresponding conservation laws (see \cite{maksjmp}) are%
\begin{equation*}
\sigma _{k,t_{n+1}}=\left( \underset{s=0}{\overset{n}{\sum }}a_{s}\sigma
_{k+n-s}\right) _{x},\text{ \ }k=1,2,...,\text{ }n=0,1,2,...
\end{equation*}

The crucial observation made in \cite{KK} is that \textit{the reduction} $%
c_{N+1}=0$ \textit{reduces hydrodynamic chain} (\ref{hydrochain}) \textit{to}
$N$ \textit{component non-diagonalisable hydrodynamic type systems, which
have infinitely many conservation laws, commuting flows and infinitely many
particular solutions}.

In these cases, (\ref{aca}) and (\ref{conslo}) reduce to the form,
respectively%
\begin{equation}
1+\overset{\infty }{\underset{m=1}{\sum }}a_{m}\lambda ^{m}=\exp \left( 
\overset{N}{\underset{m=1}{\sum }}c_{m}\lambda ^{m}\right) ,\text{ \ }1+%
\overset{\infty }{\underset{m=1}{\sum }}\sigma _{m}\lambda ^{m}=\exp \left( -%
\overset{N}{\underset{m=1}{\sum }}c_{m}\lambda ^{m}\right) .  \label{exp}
\end{equation}%
The potential function $\Phi _{(N)}$ can be found in quadratures%
\begin{equation}
d\Phi _{(N)}=\overset{N}{\underset{m=1}{\sum }}a_{m}dt_{m}.  \label{fi}
\end{equation}

\section{The Extended Hodograph Method}

\label{sec:ext}

Integration of $N$ component hydrodynamic type systems by the Extended
Hodograph Method is based on the \textit{preliminary computation} of
auxiliary $N-2$ commuting flows. In a general case, this is very complicated
task even for semi-Hamiltonian hydrodynamic type systems, because usually
commuting hydrodynamic flows can be found solving some linear system of
equations with variable coefficients in partial derivatives (for further
details, see \cite{Tsar}). However, in our case they already are extracted
from higher commuting flows (\ref{comflo}) by the same reduction $c_{N+1}=0$.

\textbf{Examples}:

\textbf{I}. If $N=2$, then $c_{3}=0$. Then we have just the two-component
hydrodynamic type system (see (\ref{raz}))%
\begin{equation}
\left( 
\begin{array}{c}
c_{1} \\ 
c_{2}%
\end{array}%
\right) _{t}=\left( 
\begin{array}{cc}
c_{1} & 1 \\ 
0 & c_{1}%
\end{array}%
\right) \left( 
\begin{array}{c}
c_{1} \\ 
c_{2}%
\end{array}%
\right) _{x}.  \label{uno}
\end{equation}

\textbf{II}. If $N=3$, then $c_{4}=0$. Then we have two three-component
commuting hydrodynamic type systems (see (\ref{raz}) again)%
\begin{equation}
\left( 
\begin{array}{c}
c_{1} \\ 
c_{2} \\ 
c_{3}%
\end{array}%
\right) _{t}=\left( 
\begin{array}{ccc}
c_{1} & 1 & 0 \\ 
0 & c_{1} & 1 \\ 
0 & 0 & c_{1}%
\end{array}%
\right) \left( 
\begin{array}{c}
c_{1} \\ 
c_{2} \\ 
c_{3}%
\end{array}%
\right) _{x},  \label{due}
\end{equation}%
\begin{equation*}
\left( 
\begin{array}{c}
c_{1} \\ 
c_{2} \\ 
c_{3}%
\end{array}%
\right) _{y}=\left( 
\begin{array}{ccc}
\frac{1}{2}c_{1}^{2}+c_{2} & c_{1} & 1 \\ 
0 & \frac{1}{2}c_{1}^{2}+c_{2} & c_{1} \\ 
0 & 0 & \frac{1}{2}c_{1}^{2}+c_{2}%
\end{array}%
\right) \left( 
\begin{array}{c}
c_{1} \\ 
c_{2} \\ 
c_{3}%
\end{array}%
\right) _{x}.
\end{equation*}

\textbf{III}. If $N=4$, then $c_{5}=0$. Then we have three four-component
commuting hydrodynamic type systems (see (\ref{raz}), (\ref{dva}))%
\begin{equation*}
\left( 
\begin{array}{c}
c_{1} \\ 
c_{2} \\ 
c_{3} \\ 
c_{4}%
\end{array}%
\right) _{t}=\left( 
\begin{array}{cccc}
a_{1} & 1 & 0 & 0 \\ 
0 & a_{1} & 1 & 0 \\ 
0 & 0 & a_{1} & 1 \\ 
0 & 0 & 0 & a_{1}%
\end{array}%
\right) \left( 
\begin{array}{c}
c_{1} \\ 
c_{2} \\ 
c_{3} \\ 
c_{4}%
\end{array}%
\right) _{x},
\end{equation*}%
\begin{equation*}
\left( 
\begin{array}{c}
c_{1} \\ 
c_{2} \\ 
c_{3} \\ 
c_{4}%
\end{array}%
\right) _{y}=\left( 
\begin{array}{cccc}
a_{2} & a_{1} & 1 & 0 \\ 
0 & a_{2} & a_{1} & 1 \\ 
0 & 0 & a_{2} & a_{1} \\ 
0 & 0 & 0 & a_{2}%
\end{array}%
\right) \left( 
\begin{array}{c}
c_{1} \\ 
c_{2} \\ 
c_{3} \\ 
c_{4}%
\end{array}%
\right) _{x},
\end{equation*}%
\begin{equation*}
\left( 
\begin{array}{c}
c_{1} \\ 
c_{2} \\ 
c_{3} \\ 
c_{4}%
\end{array}%
\right) _{z}=\left( 
\begin{array}{cccc}
a_{3} & a_{2} & a_{1} & 1 \\ 
0 & a_{3} & a_{2} & a_{1} \\ 
0 & 0 & a_{3} & a_{2} \\ 
0 & 0 & 0 & a_{3}%
\end{array}%
\right) \left( 
\begin{array}{c}
c_{1} \\ 
c_{2} \\ 
c_{3} \\ 
c_{4}%
\end{array}%
\right) _{x},
\end{equation*}%
where (we remind, see (\ref{ac})) $a_{1}=c_{1},a_{2}=c_{2}+\frac{1}{2}%
c_{1}^{2}$ and $a_{3}=c_{3}+c_{1}c_{2}+\frac{1}{6}c_{1}^{3}$.

So, now the construction of first $N-1$ commuting hydrodynamic type systems
becomes obvious. Due to a triangular form of all above square matrices, each
of them possesses just a single root (i.e. the $N$th matrix has a single
root $a_{N}$).

\subsection{The Hodograph Method}

\label{subsec:hodo}

Here we apply a classical hodograph method for integration of two-component
hydrodynamic type system (\ref{uno}). As usual such a system can be written
in the potential form%
\begin{equation*}
d\xi _{1}=c_{1}dx+\left( c_{2}+\frac{1}{2}c_{1}^{2}\right) dt,\text{ \ }d\xi
_{2}=\left( c_{2}-\frac{1}{2}c_{1}^{2}\right) dx-\frac{1}{3}c_{1}^{3}dt.
\end{equation*}%
Then one can introduce the auxiliary potentials%
\begin{equation*}
d\tilde{\xi}_{1}=xdc_{1}+td\left( c_{2}+\frac{1}{2}c_{1}^{2}\right) ,\text{
\ }d\tilde{\xi}_{2}=xd\left( c_{2}-\frac{1}{2}c_{1}^{2}\right) -\frac{1}{3}%
td(c_{1}^{3}).
\end{equation*}%
Now field variables $c_{1}$ and $c_{2}$ became new independent variables,
while independent variables $x$ and $t$ became new field variables.
Existence of the potential functions $\tilde{\xi}_{1}$ and $\tilde{\xi}_{2}$
means that their second order derivatives must coincide, i.e.%
\begin{equation}
\frac{\partial x}{\partial c_{2}}=\frac{\partial t}{\partial c_{1}}-c_{1}%
\frac{\partial t}{\partial c_{2}},\text{ \ }\frac{\partial x}{\partial c_{1}}%
=-c_{1}\frac{\partial t}{\partial c_{1}}.  \label{first}
\end{equation}%
The compatibility condition $\partial (\partial x/\partial c_{2})/\partial
c_{1}=\partial (\partial x/\partial c_{1})/\partial c_{2}$ yields the Heat
equation%
\begin{equation}
\frac{\partial t}{\partial c_{2}}=\frac{\partial ^{2}t}{\partial
c_{1}\partial c_{1}}.  \label{heat}
\end{equation}%
Once solution of this equation is found, another function $x(c_{1},c_{2})$
can be found in quadratures (see (\ref{first})), i.e.%
\begin{equation*}
d(x+c_{1}t)=tdc_{1}+\frac{\partial t}{\partial c_{1}}dc_{2}.
\end{equation*}%
Thus, here we show that two-component hydrodynamic type system (\ref{uno})
is connected with the Heat equation (\ref{heat}) by the hodograph
transformation.

\subsection{The Three Component Case}

\label{subsec:extri}

Here we apply the Extended Hodograph Method for integration of two commuting
three-component hydrodynamic type systems (\ref{due}). These systems also
can be written in the potential form%
\begin{equation*}
d\xi _{1}=c_{1}dx+\left( c_{2}+\frac{1}{2}c_{1}^{2}\right) dt+\left(
c_{3}+c_{1}c_{2}+\frac{1}{6}c_{1}^{3}\right) dy,
\end{equation*}%
\begin{equation*}
d\xi _{2}=\left( c_{2}-\frac{1}{2}c_{1}^{2}\right) dx+\left( c_{3}-\frac{1}{3%
}c_{1}^{3}\right) dt+\left( \frac{1}{2}c_{2}^{2}-\frac{1}{2}c_{1}^{2}c_{2}-%
\frac{1}{8}c_{1}^{4}\right) dy,
\end{equation*}%
\begin{equation*}
d\xi _{3}=\left( c_{3}-c_{1}c_{2}+\frac{1}{6}c_{1}^{3}\right) dx+\left( -%
\frac{1}{2}c_{2}^{2}-\frac{1}{2}c_{2}c_{1}^{2}+\frac{1}{8}c_{1}^{4}\right)
dt+\left( -c_{1}c_{2}^{2}+\frac{1}{20}c_{1}^{5}\right) dy.
\end{equation*}%
Then one can introduce the auxiliary potentials%
\begin{equation*}
d\tilde{\xi}_{1}=\left( x+tc_{1}+yc_{2}+\frac{1}{2}yc_{1}^{2}\right)
dc_{1}+(t+yc_{1})dc_{2}+ydc_{3},
\end{equation*}%
\begin{equation*}
d\tilde{\xi}_{2}=\left( -xc_{1}-tc_{1}^{2}-yc_{2}c_{1}-\frac{1}{2}%
yc_{1}^{3}\right) dc_{1}+\left( x+yc_{2}-\frac{1}{2}yc_{1}^{2}\right)
dc_{2}+tdc_{3},
\end{equation*}%
\begin{equation*}
d\tilde{\xi}_{3}=\left( -xc_{2}+\frac{1}{2}xc_{1}^{2}-tc_{2}c_{1}+\frac{1}{2}%
tc_{1}^{3}-yc_{2}^{2}+\frac{1}{4}yc_{1}^{4}\right) dc_{1}+\left(
-xc_{1}-tc_{2}-\frac{1}{2}tc_{1}^{2}-2yc_{1}c_{2}\right) dc_{2}+xdc_{3}.
\end{equation*}%
Now field variables $c_{1},c_{2}$ and $c_{3}$ became new independent
variables, while independent variables $x,t$ and $y$ became new field
variables. Existence of the potential functions $\tilde{\xi}_{1},\tilde{\xi}%
_{2}$ and $\tilde{\xi}_{3}$ means that their second order derivatives must
coincide, i.e.%
\begin{equation*}
\frac{\partial x}{\partial c_{1}}=\left( \frac{1}{2}c_{1}^{2}-c_{2}\right) 
\frac{\partial y}{\partial c_{1}},\text{ \ }\frac{\partial x}{\partial c_{2}}%
=-c_{1}\frac{\partial y}{\partial c_{1}}+\left( \frac{1}{2}%
c_{1}^{2}-c_{2}\right) \frac{\partial y}{\partial c_{2}},\text{ \ }\frac{%
\partial x}{\partial c_{3}}=\frac{\partial y}{\partial c_{1}}-c_{1}\frac{%
\partial y}{\partial c_{2}}+\left( \frac{1}{2}c_{1}^{2}-c_{2}\right) \frac{%
\partial y}{\partial c_{3}},
\end{equation*}%
\begin{equation*}
\frac{\partial t}{\partial c_{1}}=-c_{1}\frac{\partial y}{\partial c_{1}},%
\text{ \ }\frac{\partial t}{\partial c_{2}}=\frac{\partial y}{\partial c_{1}}%
-c_{1}\frac{\partial y}{\partial c_{2}},\text{ \ }\frac{\partial t}{\partial
c_{3}}=\frac{\partial y}{\partial c_{2}}-c_{1}\frac{\partial y}{\partial
c_{3}}.
\end{equation*}%
The compatibility conditions $\partial (\partial x/\partial c_{k})/\partial
c_{m}=\partial (\partial x/\partial c_{m})/\partial c_{k}$, $\partial
(\partial t/\partial c_{k})/\partial c_{m}=\partial (\partial t/\partial
c_{m})/\partial c_{k}$ yield the system%
\begin{equation*}
\frac{\partial y}{\partial c_{2}}=\frac{\partial ^{2}y}{\partial
c_{1}\partial c_{1}},\text{ \ }\frac{\partial y}{\partial c_{3}}=\frac{%
\partial ^{2}y}{\partial c_{1}\partial c_{2}},\text{ \ }\frac{\partial ^{2}y%
}{\partial c_{1}\partial c_{3}}=\frac{\partial ^{2}y}{\partial c_{2}\partial
c_{2}}.
\end{equation*}%
This means that the function $y(c_{1},c_{2},c_{3})$ satisfies simultaneously
the Heat equation and its first higher commuting flow%
\begin{equation}
\frac{\partial y}{\partial c_{2}}=\frac{\partial ^{2}y}{\partial
c_{1}\partial c_{1}},\text{ \ }\frac{\partial y}{\partial c_{3}}=\frac{%
\partial ^{3}y}{\partial c_{1}\partial c_{1}\partial c_{1}},  \label{heat2}
\end{equation}%
and the functions $t(c_{1},c_{2},c_{3})$ and $x(c_{1},c_{2},c_{3})$ can be
found in quadratures%
\begin{equation*}
d(t+yc_{1})=\frac{\partial y}{\partial c_{2}}dc_{3}+\frac{\partial y}{%
\partial c_{1}}dc_{2}+ydc_{1},
\end{equation*}%
\begin{equation*}
d\left[ x+tc_{1}+y\left( c_{2}+\frac{1}{2}c_{1}^{2}\right) \right] =\frac{%
\partial y}{\partial c_{1}}dc_{3}+ydc_{2}+(t+yc_{1})dc_{1}.
\end{equation*}%
Thus, here we show that two three-component hydrodynamic type systems (\ref%
{due}) are connected with the Heat equation and its first commuting flow (%
\ref{heat2}) by the extended hodograph transformation $c_{k}(x,t,y)%
\rightarrow x(c_{1},c_{2},c_{3}),t(c_{1},c_{2},c_{3}),y(c_{1},c_{2},c_{3})$.

\subsection{The Multi Component Case}

\label{subsec:extfo}

So, in the $N$ component case, we have $N$ time variables $t_{m}$ and $N$
field variables $c_{k}$. Taking into account identities%
\begin{equation*}
\overset{N}{\underset{m=1}{\sum }}\frac{\partial t_{k}}{\partial c_{m}}\frac{%
\partial c_{m}}{\partial t_{n}}=\delta _{k}^{n},\text{ \ }\overset{N}{%
\underset{m=1}{\sum }}\frac{\partial t_{m}}{\partial c_{n}}\frac{\partial
c_{k}}{\partial t_{m}}=\delta _{k}^{n},
\end{equation*}%
one can linearise $N$ component hydrodynamic type systems (see (\ref{comflo}%
) with the reduction $c_{N+1}=0$), i.e.

\begin{equation*}
\frac{\partial t_{k}}{\partial c_{N-n}}=\overset{N}{\underset{m=k}{\sum }}%
\sigma _{m-k}\frac{\partial t_{N}}{\partial c_{m-n}},\text{ \ }n=0,...,k-1;%
\text{ \ }\frac{\partial t_{k}}{\partial c_{N-n}}=\overset{N}{\underset{m=n+1%
}{\sum }}\sigma _{m-k}\frac{\partial t_{N}}{\partial c_{m-n}},\text{ }%
n=k,...,N-1,
\end{equation*}%
where $k=1,2,...,N-1$.

If $k=N-1$, the compatibility conditions for the sub-system%
\begin{equation*}
\frac{\partial t_{N-1}}{\partial c_{n}}=\frac{\partial t_{N}}{\partial
c_{n-1}}-c_{1}\frac{\partial t_{N}}{\partial c_{n}},\text{ \ }n=2,...,N;%
\text{ \ }\frac{\partial t_{N-1}}{\partial c_{1}}=-c_{1}\frac{\partial t_{N}%
}{\partial c_{1}}
\end{equation*}%
lead to $N-1$ commuting flows of the Heat hierarchy%
\begin{equation}
\frac{\partial t_{N}}{\partial c_{2}}=\frac{\partial ^{2}t_{N}}{\partial
c_{1}^{2}},\text{ \ }\frac{\partial t_{N}}{\partial c_{3}}=\frac{\partial
^{3}t_{N}}{\partial c_{1}^{3}},...,\text{ \ }\frac{\partial t_{N}}{\partial
c_{N}}=\frac{\partial ^{N}t_{N}}{\partial c_{1}^{N}}.  \label{hh}
\end{equation}%
Once some particular \textit{common} solution of these equations is found,
all other time variables can be found in quadratures, i.e.%
\begin{equation*}
dt_{N-n}+\overset{n}{\underset{m=1}{\sum }}a_{m}dt_{N-n+m}=\overset{N}{%
\underset{m=n+1}{\sum }}\frac{\partial t_{N}}{\partial c_{m-n}}dc_{m},\text{
\ }n=1,...,N-1.
\end{equation*}%
Moreover, the dependencies $t_{k}(\mathbf{c})$ can be expressed explicitly.
Indeed,%
\begin{equation}
t_{k}=\overset{N-k}{\underset{m=0}{\sum }}\sigma _{m}\frac{\partial U_{N}}{%
\partial c_{k+m}},\text{ \ }k=1,...,N,  \label{tk}
\end{equation}%
where the function $U_{N}$ satisfies the reduced Heat hierarchy (\ref{hh})%
\begin{equation*}
\frac{\partial U_{N}}{\partial c_{2}}=\frac{\partial ^{2}U_{N}}{\partial
c_{1}^{2}},\text{ \ }\frac{\partial U_{N}}{\partial c_{3}}=\frac{\partial
^{3}U_{N}}{\partial c_{1}^{3}},...,\text{ \ }\frac{\partial U_{N}}{\partial
c_{N}}=\frac{\partial ^{N}U_{N}}{\partial c_{1}^{N}}.
\end{equation*}%
Thus, the potential function $\Phi _{(N)}$ (see (\ref{fi})) also can be
found explicitly, i.e.%
\begin{equation}
\Phi _{(N)}=-\overset{N}{\underset{m=1}{\sum }}\sigma _{m}\frac{\partial
U_{N}}{\partial c_{m}}-U_{N}.  \label{fn}
\end{equation}

\section{The Three Dimensional Mikhal\"{e}v System}

\label{sec:mikh}

In previous papers \cite{maksjmp} and \cite{energy} diagonalisable
hydrodynamic reductions and dispersive reductions were investigated for the
Mikhal\"{e}v system (see \cite{mikh})%
\begin{equation}
a_{1,t}=a_{2,x},\text{ \ }a_{1}a_{2,x}+a_{1,y}=a_{2}a_{1,x}+a_{2,t},
\label{mikh}
\end{equation}%
whose Lax pair is%
\begin{equation}
p_{t}=[(\lambda +a_{1})p]_{x},\text{ \ }p_{y}=[(\lambda ^{2}+a_{1}\lambda
+a_{2})p]_{x}.  \label{pi}
\end{equation}%
The substitution (see (\ref{conslo}), $\lambda \rightarrow \infty $)%
\begin{equation*}
p=\exp \left( -\overset{\infty }{\underset{m=1}{\sum }}c_{m}\lambda
^{-m}\right)
\end{equation*}%
into (\ref{pi}) yields two first commuting hydrodynamic chains (\ref{raz}).

Mikhal\"{e}v system (\ref{mikh}) also can be obtained from (\ref{raz}).
Indeed, one can take first two equations from the first hydrodynamic chain%
\begin{equation*}
c_{1,t}=c_{2,x}+c_{1}c_{1,x},\text{ \ }c_{2,t}=c_{3,x}+c_{1}c_{2,x},
\end{equation*}%
and just one equation from the second hydrodynamic chain%
\begin{equation*}
c_{1,y}=c_{3,x}+c_{1}c_{2,x}+\left( c_{2}+\frac{1}{2}c_{1}^{2}\right)
c_{1,x}.
\end{equation*}%
Eliminating the field variable $c_{3}$, one can obtain the two-component
three-dimensional system%
\begin{equation*}
c_{1,t}=c_{2,x}+c_{1}c_{1,x},\text{ \ }c_{2,t}=c_{1,y}-\left( c_{2}+\frac{1}{%
2}c_{1}^{2}\right) c_{1,x},
\end{equation*}%
which is equivalent to Mikhal\"{e}v system (\ref{mikh}), where (see (\ref{ac}%
)) $a_{1}=c_{1}$, $a_{2}=c_{2}+c_{1}^{2}/2$.

Taking into account (see (\ref{chaf})) $a_{1}=\Phi _{x},a_{2}=\Phi _{t}$,
Mikhal\"{e}v system (\ref{mikh}) reduces to the Mikhal\"{e}v equation%
\begin{equation}
\Phi _{x}\Phi _{xt}+\Phi _{xy}=\Phi _{t}\Phi _{xx}+\Phi _{tt}.  \label{miha}
\end{equation}

In a general case, computation of particular solutions based on the method
of two-dimensional diagonalisable reductions is a very complicated task,
because some intermediate calculations contain integrations of linear
systems with variable coefficients in partial derivatives (for further
details, see \cite{FK}). However, in the case of non-diagonalisable
hydrodynamic reductions considered in the previous Section, corresponding
solutions of Mikhal\"{e}v equation (\ref{miha}) can be written in implicit
form.

Without loss of generality and for simplicity here we consider the
three-component case (\ref{due}) only. For better presentation below we
denote $c_{1}=X,c_{2}=T$ and $c_{3}=Y$. Now we utilise formulas (\ref{tk})
and (\ref{fn}).

\textbf{Lemma}: \textit{The particular solution of three-dimensional Mikhal%
\"{e}v equation} (\ref{miha}) \textit{selected by non-diagonalisable
hydrodynamic reduction} (\ref{due}) \textit{can be presented in the implicit
form}%
\begin{equation*}
\Phi (x,t,y)=\left( Y-XT+\frac{1}{6}X^{3}\right) \frac{\partial U}{\partial Y%
}+\left( T-\frac{1}{2}X^{2}\right) \frac{\partial U}{\partial T}+X\frac{%
\partial U}{\partial X}-U,
\end{equation*}%
\textit{where the function} $U(X,T,Y)$ \textit{is a solution of the pair of
commuting Heat equations}%
\begin{equation}
\frac{\partial U}{\partial T}=\frac{\partial ^{2}U}{\partial X^{2}},\text{ \ 
}\frac{\partial U}{\partial Y}=\frac{\partial ^{3}U}{\partial X^{3}},
\label{hit}
\end{equation}%
\textit{and the functions }$X(x,t,y),T(x,t,y),Y(x,t,y)$ \textit{can be found
from}%
\begin{equation*}
y=\frac{\partial U}{\partial Y},\text{ \ }t=\frac{\partial U}{\partial T}-X%
\frac{\partial U}{\partial Y},\text{ \ }x=\frac{\partial U}{\partial X}-X%
\frac{\partial U}{\partial T}+\left( \frac{1}{2}X^{2}-T\right) \frac{%
\partial U}{\partial Y}.
\end{equation*}

\textbf{Example I}: If ($\kappa $ is an arbitrary constant)%
\begin{equation*}
U=\exp \left( \kappa X+\kappa ^{2}T+\kappa ^{3}Y\right) ,
\end{equation*}%
then Mikhal\"{e}v equation (\ref{miha}) has a particular solution%
\begin{equation*}
\Phi (x,t,y)=\frac{y}{\kappa ^{3}}\ln \frac{y}{\kappa ^{3}}-\frac{11}{6}%
\frac{y}{\kappa ^{3}}+\frac{t}{\kappa ^{2}}+\frac{1}{\kappa }\left( x-\frac{%
t^{2}}{2y}\right) +\frac{t^{3}}{3y^{2}}-\frac{xt}{y},
\end{equation*}%
where%
\begin{equation*}
X=\kappa ^{-1}-\frac{t}{y},\text{ \ }T=\frac{1}{2}\kappa ^{-2}-\frac{x}{y}+%
\frac{t^{2}}{2y^{2}},\text{ \ }Y=\kappa ^{-3}\left( \ln \frac{y}{\kappa ^{3}}%
-\frac{3}{2}\right) +\kappa ^{-2}\frac{t}{y}+\kappa ^{-1}\left( \frac{x}{y}-%
\frac{t^{2}}{2y^{2}}\right)
\end{equation*}%
and $U=y\kappa ^{-3}$.

\textbf{Remark}: While the Extended Hodograph Method admits linearisation of
sets of $N-1$ commuting hydrodynamic flows, the Tsarev Generalised Hodograph
Method allows to extract infinitely many particular solutions from higher
commuting dispersive chains (\ref{chaf}). For instance, non-diagonalisable
commuting hydrodynamic type systems (\ref{due}) have infinitely many
hydrodynamic conservation laws%
\begin{equation*}
\left( \frac{\partial F}{\partial c_{2}}\right) _{t}=\left( c_{1}\frac{%
\partial F}{\partial c_{2}}+\frac{\partial F}{\partial c_{1}}\right) _{x},%
\text{ \ }\left( \frac{\partial F}{\partial c_{2}}\right) _{y}=\left[ \left(
c_{2}+\frac{1}{2}c_{1}^{2}\right) \frac{\partial F}{\partial c_{2}}+c_{1}%
\frac{\partial F}{\partial c_{1}}-F\right] _{x},
\end{equation*}%
where the function $F(c_{1},c_{2},c_{3})$ satisfies (cf. (\ref{heat2}))%
\begin{equation*}
\frac{\partial F}{\partial c_{2}}=-\frac{\partial ^{2}F}{\partial
c_{1}\partial c_{1}},\text{ \ }\frac{\partial F}{\partial c_{3}}=\frac{%
\partial ^{3}F}{\partial c_{1}\partial c_{1}\partial c_{1}},
\end{equation*}%
and infinitely many higher commuting flows%
\begin{equation*}
\left( 
\begin{array}{c}
c_{1} \\ 
c_{2} \\ 
c_{3}%
\end{array}%
\right) _{\tau }=\left( 
\begin{array}{ccc}
U & \frac{\partial U}{\partial c_{1}} & \frac{\partial U}{\partial c_{2}} \\ 
0 & U & \frac{\partial U}{\partial c_{1}} \\ 
0 & 0 & U%
\end{array}%
\right) \left( 
\begin{array}{c}
c_{1} \\ 
c_{2} \\ 
c_{3}%
\end{array}%
\right) _{x},
\end{equation*}%
where the function $U(X,T,Y)$ satisfies (\ref{hit}).

\textbf{Statement}: The Heat hierarchy%
\begin{equation*}
\frac{\partial U}{\partial c_{2}}=\frac{\partial ^{2}U}{\partial
c_{1}\partial c_{1}},\text{ \ }\frac{\partial U}{\partial c_{3}}=\frac{%
\partial ^{3}U}{\partial c_{1}\partial c_{1}\partial c_{1}},\text{ \ }\frac{%
\partial U}{\partial c_{4}}=\frac{\partial ^{4}U}{\partial c_{1}\partial
c_{1}\partial c_{1}\partial c_{1}},...
\end{equation*}%
has infinitely many polynomial solutions $U_{k}(\mathbf{c})$ which are Bell
polynomials, determined by (see (\ref{aca}) and (\ref{conslo}))%
\begin{equation*}
1+\overset{\infty }{\underset{m=1}{\sum }}U_{m}\lambda ^{m}=\exp \left( 
\overset{\infty }{\underset{m=1}{\sum }}c_{m}\lambda ^{m}\right) .
\end{equation*}

\textbf{Example II}: A general solution of Mikhal\"{e}v equation (\ref{miha}%
) is determined by (see (\ref{fi}), (\ref{tk}), (\ref{fn}))%
\begin{equation*}
\Phi =c_{1}\frac{\partial U}{\partial c_{1}}+\left( c_{2}-\frac{1}{2}%
c_{1}^{2}\right) \frac{\partial U}{\partial c_{2}}+\left( c_{3}-c_{1}c_{2}+%
\frac{1}{6}c_{1}^{3}\right) \frac{\partial U}{\partial c_{3}}-U,
\end{equation*}%
where (we remind that the function $U(X,T,Y)$ satisfies (\ref{hit}))%
\begin{equation*}
y=\frac{\partial U}{\partial c_{3}},\text{ \ }t=\frac{\partial U}{\partial
c_{2}}-c_{1}\frac{\partial U}{\partial c_{3}},\text{ \ }x=\frac{\partial U}{%
\partial c_{1}}-c_{1}\frac{\partial U}{\partial c_{2}}+\left( -c_{2}+\frac{1%
}{2}c_{1}^{2}\right) \frac{\partial U}{\partial c_{3}}.
\end{equation*}%
So, infinitely many polynomial solutions $\tilde{U}_{k}(\mathbf{c})$ can be
extracted from the expansion (cf. (\ref{exp}))%
\begin{equation*}
1+\overset{\infty }{\underset{m=1}{\sum }}\tilde{U}_{m}\lambda ^{m}=\exp
\left( \overset{3}{\underset{m=1}{\sum }}c_{m}\lambda ^{m}\right) .
\end{equation*}%
For instance (here we again use the notation $c_{1}=X,c_{2}=T,c_{3}=Y$),%
\begin{equation*}
\tilde{U}_{1}=X,\text{ \ }\tilde{U}_{2}=T+\frac{1}{2}X^{2},\text{ \ }\tilde{U%
}_{3}=Y+XT+\frac{1}{6}X^{3},
\end{equation*}%
\begin{equation*}
\tilde{U}_{4}=XY+\frac{1}{2}T^{2}+\frac{1}{2}X^{2}T+\frac{1}{24}X^{4},\text{
\ }\tilde{U}_{5}=TY+\frac{1}{2}X^{2}Y+\frac{1}{2}XT^{2}+\frac{1}{6}X^{3}T+%
\frac{1}{120}X^{5}.
\end{equation*}%
Corresponding solutions of Mikhal\"{e}v equation (\ref{miha}) become ($%
k=4,5,...$)%
\begin{equation*}
\Phi _{(k)}=X\tilde{U}_{k-1}+\left( T-\frac{1}{2}X^{2}\right) \tilde{U}%
_{k-2}+\left( Y-XT+\frac{1}{6}X^{3}\right) \tilde{U}_{k-3}-\tilde{U}_{k},
\end{equation*}%
where%
\begin{equation*}
x=\tilde{U}_{k-1}-X\tilde{U}_{k-2}+\left( -T+\frac{1}{2}X^{2}\right) \tilde{U%
}_{k-3},\text{ \ }t=\tilde{U}_{k-2}-X\tilde{U}_{k-3},\text{ \ }y=\tilde{U}%
_{k-3}.
\end{equation*}%
In the simplest nontrivial case $\tilde{U}_{4}$, we have three commuting
hydrodynamic type system (cf. (\ref{due}))%
\begin{equation*}
\left( 
\begin{array}{c}
c_{1} \\ 
c_{2} \\ 
c_{3}%
\end{array}%
\right) _{t}=\left( 
\begin{array}{ccc}
\tilde{U}_{1} & 1 & 0 \\ 
0 & \tilde{U}_{1} & 1 \\ 
0 & 0 & \tilde{U}_{1}%
\end{array}%
\right) \left( 
\begin{array}{c}
c_{1} \\ 
c_{2} \\ 
c_{3}%
\end{array}%
\right) _{x},
\end{equation*}%
\begin{equation*}
\left( 
\begin{array}{c}
c_{1} \\ 
c_{2} \\ 
c_{3}%
\end{array}%
\right) _{y}=\left( 
\begin{array}{ccc}
\tilde{U}_{2} & \tilde{U}_{1} & 1 \\ 
0 & \tilde{U}_{2} & \tilde{U}_{1} \\ 
0 & 0 & \tilde{U}_{2}%
\end{array}%
\right) \left( 
\begin{array}{c}
c_{1} \\ 
c_{2} \\ 
c_{3}%
\end{array}%
\right) _{x}.
\end{equation*}%
\begin{equation*}
\left( 
\begin{array}{c}
c_{1} \\ 
c_{2} \\ 
c_{3}%
\end{array}%
\right) _{z}=\left( 
\begin{array}{ccc}
\tilde{U}_{3} & \tilde{U}_{2} & \tilde{U}_{1} \\ 
0 & \tilde{U}_{3} & \tilde{U}_{2} \\ 
0 & 0 & \tilde{U}_{3}%
\end{array}%
\right) \left( 
\begin{array}{c}
c_{1} \\ 
c_{2} \\ 
c_{3}%
\end{array}%
\right) _{x}.
\end{equation*}%
According to \cite{GK}, \cite{Tsar} one can obtain the system\footnote{%
The Tsarev Generalised Hodograph Method reformulated by Yu. Kodama and J.
Gibbons for a three-component case means: $x\delta
_{k}^{i}+tU_{k}^{i}+yV_{k}^{i}=W_{k}^{i}$, where $U_{k}^{i}$ and $V_{k}^{i}$
are velocity matrices of two commuting hydrodynamic type systems, while $%
W_{k}^{i}$ is a velocity matrix of any other commuting higher commuting
flow. $\delta _{k}^{i}$ is the Kronecker delta.}%
\begin{equation*}
\tilde{U}_{3}=x+t\tilde{U}_{1}+y\tilde{U}_{2},\text{ \ }\tilde{U}_{2}=t+y%
\tilde{U}_{1},\text{ \ }\tilde{U}_{1}=y.
\end{equation*}%
Then (see (\ref{fi}), (\ref{tk}), (\ref{fn}))%
\begin{equation*}
d\Phi =\tilde{U}_{1}dx+\tilde{U}_{2}dt+\tilde{U}_{3}dy
\end{equation*}%
can be integrated explicitly. A corresponding solution of Mikhal\"{e}v
equation (\ref{miha}) is%
\begin{equation*}
\Phi =xy+\frac{1}{2}t^{2}+ty^{2}+\frac{1}{4}y^{4}.
\end{equation*}%
The next case $\tilde{U}_{5}$ leads to the non-polynomial solution of Mikhal%
\"{e}v equation (\ref{miha})%
\begin{equation*}
\Phi =yt+\frac{1}{5}X^{5},
\end{equation*}%
where (here $Y=t+\frac{1}{3}X^{3},T=y-\frac{1}{2}X^{2}$)%
\begin{equation*}
X^{4}=4x+2y^{2}.
\end{equation*}

So, the existence of non-diagonalisable hydrodynamic reductions for
integrable three-dimensional quasilinear systems of first order is also a
new interesting phenomenon. Thus, examination of multi-dimensional
quasilinear systems of first order should be continued by this reason in
further publications.

\section{Conclusion}

\label{sec:final}

Strictly hyperbolic hydrodynamic type systems have pairwise distinct roots.
In this paper we considered the opposite situation: hydrodynamic type
systems with multiple roots. The most degenerate case, just a one multiple
root. This means that such hydrodynamic type systems no longer are
hyperbolic. They are parabolic. The transformation to the Heat equation
hierarchy is a natural illustration of this parabolic phenomenon.

An effective mechanism of integration of hydrodynamic type systems with
pairwise distinct roots based on existence of special coordinate system,
i.e. the Riemann invariants. In the case considered in this paper the
coordinate system $c_{k}$ is a most convenient from computational point of
view. These coordinates have a natural interpretation, in these coordinates
the Heat hierarchy is written as an infinite set of commuting linear
equations in partial derivatives with constant coefficients.

A straightforward computation shows that, for instance, the Benney
hydrodynamic chain does not possess such non-diagonalisable hydrodynamic
reductions (hydrodynamic type systems with multiple roots). So, we come to
the observation: just linearly degenerate hydrodynamic chains (they have at
least one $N$ component linearly degenerate hydrodynamic reduction for each $%
N$, for further details see \cite{fer91}, \cite{fermos}, \cite{makslin})
possess hydrodynamic reductions with multiple roots. Inspection of other
linearly degenerate hydrodynamic chains should be made in a separate paper.

\section*{Acknowledgements}

The author is grateful to Yuri Brezhnev, Alexander Chupakhin, Eugeni
Ferapontov, Gennady El, Yuji Kodama and Boris Konopelchenko for many useful
discussions. This work was partially supported by the RFBR grant No.
16-51-55012. The author also appreciates kind hospitality at the Tsinghua
University (Beijing, China), where the main part of this research was
completed.

\end{document}